# Information retrieval from a phoneme time series database

Radhakrishnan Nagarajan
University of Arkansas for Medical Sciences, Little Rock, AR 72205
*rnagarajan@uams.edu*
Anand Nagarajan
Symbram LLC, Sunnyvale, CA
*anand@esymbram.com*
Mariofanna Milanova
University of Arkansas at Little Rock, Little Rock, AR 72204
*mgmilanova@ualr.edu*

**Abstract**
*Developing fast and efficient algorithms for retrieval of objects to a given user query is an area of active research. The present study investigates retrieval of time series objects from a phoneme database to a given user pattern or query. The proposed method maps the one-dimensional time series retrieval into a sequence retrieval problem by partitioning the multi-dimensional phase-space using k-means clustering. The problem of whole sequence as well as subsequence matching is considered. Robustness of the proposed technique is investigated on phoneme time series corrupted with additive white Gaussian noise. The shortcoming of classical power-spectral techniques for time series retrieval is also discussed.*

## 1. Introduction

Real-world physical and biological systems are inherently nonlinear feedback systems that evolve with time. External recording from such a system results in data dicretized in *time* as well as *amplitude* also know as *digital* data. Such a digital data generated as function of time represents a *time series*. Time series analysis is quite an involved topic accompanied by rigorous theoretical constructs [1]. However, there has been recent emphasis on developing *generic* techniques for *approximate* time series and sequence matching from information retrieval perspective[2-5].

The present study explores a nonlinear dynamical approach for time series matching and retrieval. The interdisciplinary nature of the nonlinear dynamical systems lends itself to a wide range of applications [6, 7]. It is well appreciated that the various components of the experimental systems in real world are coupled nonlinearly with complex feedback loops. This restricts the choice of linear systems approaches for their analysis. External recordings sampled from such systems provide insight into the underlying dynamics and may exist in *more* than one-dimension. This multi dimensional representation is also known as the *state-space/phase-space* representation of the time series. The theory of nonlinear dynamical analysis provides us a way to reconstruct the multi-dimensional representation from the observed one-dimensional realization. Such a multi-dimensional representation also captures the rich geometry and inherent nonlinear correlations not evident in the one-dimensional representation. The report is organized as follows: In Sec 2, the mapping on the one-dimensional time series onto a multi-dimensional phase-space is discussed. Subsequently, partitioning of the vectors in the phase space into a binary sequence is discussed in Sec 3. Such an approach implicitly maps the time series matching problem into a sequence matching problem. In Sec 4, whole sequence and subsequence retrieval from a phoneme database to a given user query is discussed. The effectiveness of the proposed technique in the presence of additive white Gaussian noise (AWGN) is also investigated.

## 2. Phase-Space Representation

Consider a dynamical system in a D-dimensional state space, $\mathbf{R}^D$. While the original state space may be high dimensional, in the case of *dissipative* dynamical systems, which is characteristic of most real world systems, the dynamics settles down on a low dimensional *attractor* with system evolution. This attractor exhibits rich geometry. The one-dimensional time series of finite number of points ($N$) is obtained by sampling a *single* dynamical variable with the *measurement function* $\varphi$ at finite intervals of time ($T$), given by $\varphi_n \in \mathbf{R}$. The method of delays by Takens [8], provides us a way to reconstruct the vector series in an equivalent state space in $\mathbf{R}^d$ from the observed one-dimensional time series $\varphi_n$, $n = 1...N$. Such a mapping has been found to preserve the topological properties of the original state space. The reconstructed vector series in the equivalent state space is given by

$$w(k)=(\varphi_n\ \varphi_{n+t}\ \varphi_{n+2t}\ ...\ \varphi_{n+(d-1)\tau}),\ 1 \le k \le N - (d-1)\tau\ ..(1)$$

where $w(k)$ represents the state of the system in the $d$-dimensional space. The above procedure is called *embedding* [8]. The quantities $d$ and $\tau$ in (1) represent the *embedding dimension* and the *time delay* respectively. The embedded vectors can be represented in a matrix form, known as the *trajectory matrix* ($\Gamma$), where

$$\Gamma = [w(1), w(2), ..., w(N-(d-1)\tau)]^T \quad \text{...... (2)}$$

Recent extensions and modifications to the Taken's theorem have increased the application of the concepts of nonlinear dynamics to a wide range of data sets. Stark et al., [8] extended the Taken's theorem to the class of forced systems. The concept of over-embedding was recently suggested [9] even to accommodate nonstationarities due to slowly varying time dependent parameters.

*Example 1*: Consider the one-dimensional time series of real numbers ($\varphi_1\ \varphi_2\ \varphi_3\ \varphi_4\ \varphi_5\ \varphi_6\ \varphi_7\ \varphi_8$), N = 8 samples. For $d = 3$, $\tau = 1$, we obtain N - (d-1).$\tau$ = 6 vectors in the 3-dimensional phase-space given by

$w(1) = (\varphi_1\ \varphi_2\ \varphi_3); w(2) = (\varphi_2\ \varphi_3\ \varphi_4); w(3) = (\varphi_3\ \varphi_4\ \varphi_5);$
$w(4) = (\varphi_4\ \varphi_5\ \varphi_6); w(5) = (\varphi_5\ \varphi_6\ \varphi_7); w(6) = (\varphi_6\ \varphi_7\ \varphi_8)$

The vectors form the rows of the trajectory matrix ($\Gamma$). It should be noted that a proper choice of the embedding dimension ($d$) and time delay ($\tau$) is necessary for a proper unfolding of the geometry in the phase space.

### 2.1. Optimal embedding parameters (d, τ)

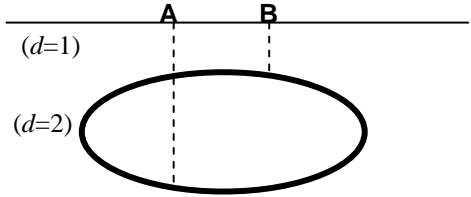

**Fig. 1. Two points (A, B) at diametrically opposite ends of an ellipse ($d$ = 2) tend to be false nearest neighbors when projected on a line ($d$=1). Unlike ($d$ = 2), ($d$ = 1) fails to capture the true geometry of the points.**

If $d_e$ is the true dimension of the data in the original state-space then $d > 2d_e$ would be a *sufficient* choice of the *embedding dimension* [6,7]. In real world data sets one does not have any knowledge about $d_e$. In such cases, the method of false-nearest neighbor's (FNN) [11] is used to determine the minimal embedding dimension ($d$). The vectors that are close to each other in a lower dimension, may show significant separation as the dimension is increased. The fraction of the FNN decreases with the increase in the embedding dimension. The value at which FNN is lowest (almost zero) is the minimal sufficient embedding dimension. A popular example where tow close to each other in a one-dimensional representation show significant separation while embedded onto two-dimensions, Fig. 1, to elucidate this point. The embedding dimension estimated for a few of the phoneme data sets by the FNN technique is shown in Fig 2. It can be observed that the number of FNN is almost zero for embedding dimension, d ~ 4. Several techniques have been proposed for the choice of the time delay $\tau$ [12, 13]. In this report, the time delay is chosen as ($\tau$ = 1).

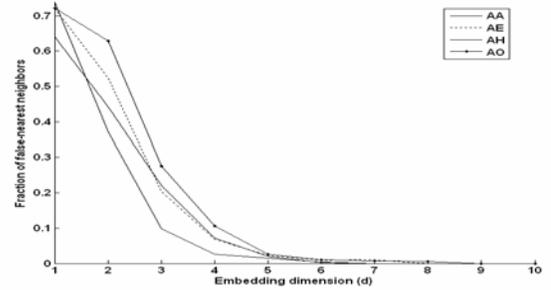

**Fig. 2. Fraction of the false nearest neighbors (FNN) as a function of the embedding dimension ($d$) for phoneme data sets *AA, AE, AH*, and *AO*. The number of FNN decreases with increase in the embedding dimension ($d$) characteristic of dissipative dynamical systems.**

### 2.3 Power-spectral analysis - shortcomings

Linear transformation techniques such as Power spectral analysis, singular value decomposition (PCA), have been used extensively in mining time series data [3, 4]. A close connection exists between the embedding procedure and harmonic decomposition of periodic signals [14]. Consider a linear process generated by $m$ harmonically related sinusoids with distinct frequencies. The singular value decomposition (SVD) of the trajectory matrix ($\Gamma$) constructed with, $\tau = 1$ and $d = m+1$, is sufficient to determine the dominant frequency components, and hence the spectral content of the data. A sharp decrease in magnitude is observed between the $m^{th}$ and the $(m+1)^{th}$ eigen-value. This approach is popularly known as *Pisarenko's harmonic decomposition* [15, 16] and relies on *Caratheodory's uniqueness result* [17]. Related techniques are Multiple SIgnal Classification (MUSIC) and Estimation of Signal Parameters via Rotational Invariance Technique (ESPRIT) [16]. These techniques implicitly assume that the vectors in $\Gamma$ are linearly correlated, which is not true for nonlinearly correlated data. Therefore, SVD of $\Gamma$, results in spurious eigen-values.

To further illustrate the point that two data sets can have the *same spectral signature but different geometry*. This is illustrated by comparing the phase-space representation of the phoneme AA to those of its *phase-randomized surrogate* (AA*) [18].

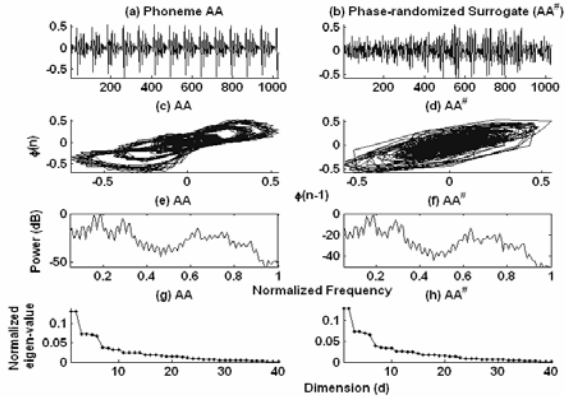

**Fig. 3.** Phoneme AH and its phase-randomized surrogate (AH#) are shown in (a) and (b) respectively. Their geometry upon embedding (*d* = 2, $\tau$ = 1) is markedly different and shown in (c) and (d) respectively. However, their power-spectrum exhibit considerable similarity across a wide-range of frequencies and shown in (e) and (f) respectively. Similarity in their power-spectrum is also reflected by the similarity in their normalized eigen-value spectra obtained by SVD of the trajectory matrix with (*d* = 30, $\tau$ = 1), shown in (g) and (h) respectively.

Phase-randomized surrogate are constrained bootstrap realization generated so as to retain the power-spectrum of the given data [17, 18]. Retaining the power-spectrum implicitly results in retaining the auto-correlation function (Wiener-Khinchine theorem) [19] and the optimal parameters of a linearly correlated Gaussian noise are implicitly determined from the auto-correlation function (Yule-Walker equations) [19]. In a recent study [3] a subset of the spectral components were used to successfully retrieve time series objects to a user query. However, it is important to note that power-spectral approaches *may not be sufficient* to completely the dynamics in the presence of non-Gaussianity and nonlinearity. From Fig. 6, it is clear that while the spectral content of the phoneme data and its surrogate realization is similar their dynamics in the phase space are quite different. Thus it is possible that linear transformation approaches might be inadequate in explaining the dynamics inherent in phoneme sets. In our simulation results, Sec. 5, the database was populated purposely with phase-randomized surrogates of the phoneme data sets. This is done in order to establish the fact the proposed technique is sensitive to the geometry, hence can be superior to power-spectral estimation techniques.

## 3. Sequence representation

Symbolic sequence reconstruction generated from time series have been found to capture the *dynamics*, hence the temporal structure in the data [20]. Let *(W, T)* be a dynamical system in the real space where, *W* represents the set of all possible states of the system and *T* the transformation between any two adjacent states separated by one time instant. Each point in *(W, T)* can be mapped into a unique sequence in the symbol space *(Y,$\sigma$)* where *Y* represents the sequence set corresponding to the points $w \in W$ and $\sigma$ the *shift transformation* which mimics the transformation *T*. Consider the trajectory of a point $w \in W$, given by … $T^{-1}(w)$, *w*, $T(w)$, $T^2(w)$…., and a binary partitioning of *W* into two cells, Fig. 4.

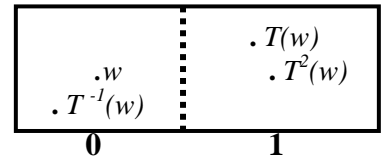

**Fig. 4.** Binary partitioning (dotted lines) of the orbit generated by *w*.

The sequence representation of $w \in W$ is given by $s(w) =$ …00.11... Marston Morse [21] showed that under certain generic conditions, a *binary sequence* representation could produce a faithful representation of the underlying dynamics. The ad-hoc approach is to obtain the sequence representation by partitioning the one-dimensional time series about a threshold point ($\theta$) such as the mean, median or mid-point [22]. i.e. $s_i = 0$, if $\varphi_i > \theta$ otherwise $s_i = 1$. This generates a binary partition with alphabet set $\Sigma = \{0, 1\}$, $n(\Sigma) = 2$. Determining such a threshold is a non-trivial issue. In a previous work, we proposed the k-means clustering technique [23] as an alternate effective method to partition the vector series in a *d-dimensional* state space with ($d \geq 1$), into a sequence with alphabet set $n(\Sigma) \geq 2$. Each of the clusters is assigned a unique symbol and the number of clusters (symbols) forms the alphabet set of the resulting sequence [23].

## 4. Sequence retrieval
### 4.1 Whole sequence retrieval

Let the user query and the *m* time series objects residing in the database be represented by $q^*$ and $s_1^*, s_2^*, ...., s_m^*$ respectively. For whole sequence retrieval the length of the query $q^*$ is equal to the length of $s_k^* \forall i = 1...m$. The corresponding sequence representations $q$ and $s_k^* \forall i = 1...m$ are generated as follows:

- Determine the embedding dimension ($d$) and delay ($\tau$), Sec. 2, for the query $q^*$.
- Generate the vector series representations $w_{q^*}$ and $w_{s_k^*}$ of $q^*$ and $s_k^*$ with these parameters ($d$, $\tau$), Sec. 2.
- Generate the binary sequence representation $q$ of $q^*$ by k-means partitioning $w_{q^*}$ with n($\Sigma$) = 2 and store the centroids in $\mathbf{z}(q^*)$. The binary sequence representation $s_k$ of $\forall k = 1...m$ are generated by comparing the value of their vector series $w_{s_k^*}$ to $\mathbf{z}(q^*)$ independently.

**Example 2**: *Consider object $s_k^*$ and query $q^*$. First partition $q^*$ into a binary sequence $q$ using k-means. Let the resulting centroids be $z_1$ and $z_2$. Determine $d_i^1 = \left\| w_{s_k^*}(i) - z_1 \right\|$ and $d_i^2 = \left\| w_{s_k^*}(i) - z_2 \right\|$ for $i = 1...length(w_{s_k^*})$.*

$$w_{s_k^*}(i) = 0 \;\; if \;\; d_i^1 < d_i^2$$
$$= 1 \;\; otherwise$$

*Repeat the above procedure for every time series object (k) in the database.*

It is important to note that *only the query needs to be clustered and not the objects residing in the database*. Since k-means partitioning is sensitive to the initial choice of the centroids [24], generating the sequence representation $s_k$ from the centroids $\mathbf{z}(q^*)$ eliminates this problem. The above procedure maps the query and the objects into sequences. In order to retrieve objects matching the given query, we use the hamming distance ($h$) in the present study. $h(q,s) = 0$ implies that the symbolic representations match bit-wise (*hard decision*). For practical considerations it might be prudent to set an error bound, such that $h(q^*, s^*) < \in$ (*soft decision*). Soft decision can prove to be meaningful in the presence of noise which is ubiquitous in real-world experimental data.

### 4.2 Subsequence retrieval

Let the user query and the $m$ time series objects residing in the database be represented by $q^*$ and $s_1^*, s_2^*, ...., s_m^*$ respectively. Unlike whole sequence retrieval, the length of $q^*$ *need not be equal* to the length of $s_k^* \forall k = 1...m$ in subsequence retrieval. This arises when the objective is to retrieve a time series with a particular pattern/query interspersed in it. A stepwise procedure is described below for generating the sequence representations of the query $q^*$ and the objects $s_k^* \forall k = 1...m$.

- The embedding dimension ($d$) and delay ($\tau$) are determined for $s_k^* \forall k = 1...m$ as described in Sec 2. Generate the vector series of $s_k^*$ with parameters ($d$, $\tau$). Generate a binary partition of the phase-space of $s_k^*$ using the k-means technique and store the centroids in $\mathbf{z}(s_k^*)$ as well as the embedding parameters $d$, and $\tau$ for any future queries. This would be equivalent to maintaining a *codebook* for each object in the database. Generate the corresponding symbolic representation $s_k$, by assigning a unique symbol to each of the clusters.
- The symbolic representation $q$ corresponding to the query $q^*$ is generated by partitioning its phase space using $\mathbf{z}(s_k^*)$ (as in Example 2). The embedding parameters $d$, $\tau$ along with the centroids $\mathbf{z}(s_k^*)$ of *that* $s_k^*$ is used directly from the codebook.
- The centroids $\mathbf{z}(s_k^*)$ and the embedding parameters in the codebook are retained for future queries.

Similar to the case of whole sequence matching, using centroid $\mathbf{z}(s_k^*)$ to cluster the query $q^*$ eliminates sensitivity of the k-means to the choice of the centroids. It is important to note that the clustering and embedding parameters of the objects $s_k^* \forall k = 1...m$ have to be determined *only once*. The embedding dimension ($d$), timed delay ($\tau$) and centroids $\mathbf{z}(s_k^*)$ of each $s_k^*$ are stored in the codebook for all future queries. Fast exact sequence searching algorithms are ideal for matching noise free sequences [25]. However, in the presence of noise, an exact match is highly unlikely. Ad-hoc approach would be to choose a sliding window with a window length equal to the length of the query and compute the hamming distance between the windowed sequence and the given query. The desired match is chosen as the window with the minimum hamming distance. However, this is undesirable from a computational stand point. Recent studies have proposed approximate sequence matching techniques [26, 27]. In this report, *agrep* algorithm (http://www.tgries.de/agrep/) was used for subsequence matching, Sec 5.

## 5. Results

The whole sequence and subsequence retrieval were tested on a database of phoneme data sets in *.au* format, encoded by 8-bit *μ-law* quantizer. These data sets are available publicly and can be obtained from the following hyperlinks: *http://www.ibiblio.org/sounds/phonemes/* (or) *http://www.speech.cs.cmu.edu/comp.speech/Section1/speechlinks.html*. The 34 phonemes considered include: *OW, UW, AH, AO, AW, AX, AY, DH, ER, EY, HH, IH, IY, NG, **AE**, OY, SH, TH, UH, **AA**, WH, YU, d, g, j, l, m, n, r, s, v, w, y, z*. These data sets were sufficiently longer (> 1024) and shall be referred by the numbers (1…34) in the subsequent discussion. The phonemes AE and AA correspond to numbers 15 and 20 respectively and used as queries in the subsequent analysis. Noisy versions were generated by adding zero mean, unit variance additive white Gaussian ($\epsilon_t$) at noise level $f \in [0,1]$ noise as $y_t = s_t + f \cdot \epsilon_t$. The corresponding signal to noise ratio is given by SNR (dB) = $20 * \log_{10}(\sigma_{signal}/\sigma_{noise})$, where $\sigma_{signal}$ and $\sigma_{noise}$ represent the standard deviation of the signal ($s_t$) and noise ($f \cdot \epsilon_t$) respectively.

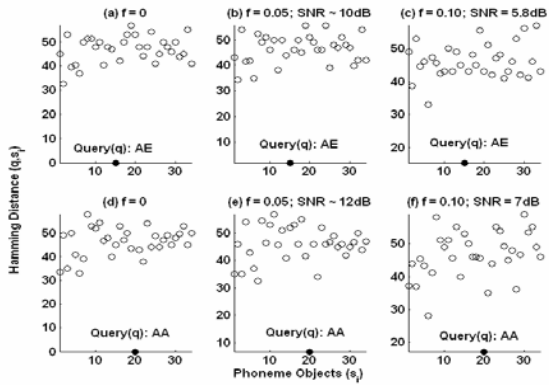

**Fig.5. The mean hamming distance (hollow circles) across 50 independent realizations across the 34 phoneme objects for query AA (401:501) as function of the noise level *f* = (0, 0.05, 0.10) is shown in (a, b and c) respectively. A similar analysis for query AE (401:501) is shown in (d, e and f) respectively. A minimum value is obtained for the matching objects (dark circles).**

### 5.1 Whole sequence retrieval

For whole sequence retrieval we considered the 34 phoneme objects and fixed their length and those of the query to 100, i.e. from (401:500). The embedding parameters were fixed as ($d = 4$, $\tau = 1$) Fig.2, for the query (Sec. 4.1). These were also retained across the 34 objects. The mean hamming distance across the 34 phoneme objects in the database for the queries **AA** (401:501) and **AE** (401:501) as function of the noise level *f* = (0, 0.05, 0.10) is shown in Figs. 5(a, b, c) and 5(d, e, f) respectively. The results obtained on the phase-randomized surrogates of the 34 phoneme data which share the same power-spectra failed to retrieve the matches as expected. *Thus the proposed technique is powerful in discerning data sets of similar correlation properties but different geometry.*

### 5.2 Sub-sequence retrieval

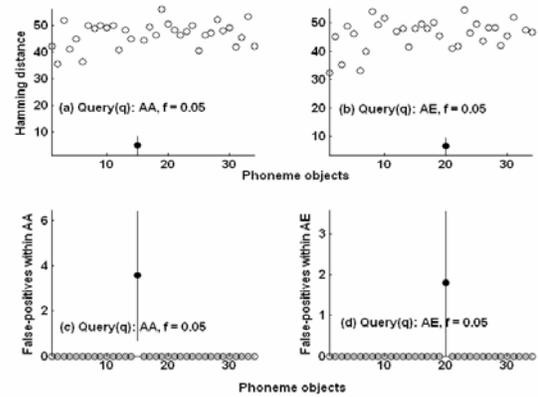

**Fig. 6. Average hamming distance estimates for 10 independent realizations of the queries AA (401:500) and AE (401:500) is shown in (a) and (b) respectively (hollow circles). These in turn we used as initial guess of the penalty (*k*) of the agrep. The number of false-positives across the 34 phoneme objects for the queries AA (401:500) and AE (401:500) is shown in (c) and (d) respectively. Matches of interest are shown by solid circles.**

For subsequence retrieval we considered the 34 phoneme data sets of arbitrary length. The embedding parameters were fixed as ($d = 4$, $\tau = 1$) for each of the objects in the database (Sec. 4.2). Binary partition of phase space across each of these data sets were generated as discussed in (Sec. 4.2). Subsequently, these centroids were stored in a codebook and used to generate the corresponding binary sequence representation of the given query. Unlike whole sequence retrieval, the length of the query is significantly shorter than the objects in the database in the case of subsequence retrieval. A naïve approach would be to use a sliding window approach and choose the window(s) that have hamming distance below a certain threshold (θ). In the present study, we use the approximate sequence matching algorithm [26, 27] *agrep* (http://www.tgries.de/agrep/) for subsequence retrieval. The penalty (*k*) incorporating deletions or insertions or gap in the agrep algorithm is user specified. In the case of noise free sequence ($k = 0$). A non-zero $k$ is essential for noisy subsequence retrieval. Rather than choose an arbitrary $k$, we encourage determining the hamming distance of noisy versions of the given query as initial guess of penalty ($k$). Noisy versions of the queries AA (401:500) and AE (401:500) at $f = 0.05$ were generated. The mean and standard deviation across 10 independent realizations is shown in Figs. 6a and 6b respectively. These values are close to 10 and conform to our earlier observations Figs. 5b and 5e. Thus we used ($k = 10$) in order to accommodate noise factor ($f = 0.05$). Agrep correctly retrieved the subsequences from the phoneme database. However, there were instances where more than one match was retrieved within the sequence of interest. It is important to note that the false-positives *were only within the sequence of interest* and not across sequences (see Figs. 6c and 6d). The mean and standard deviation of the false-positives across the 34 phoneme objects for queries AA (401:500) and AE (401:500) at $f = 0.05$ is shown in Figs. 6c and 6d respectively.

Preliminary results presented in this study clearly demonstrate that a simple binary partition of the phase-space representation of the phonemes can be useful for their retrieval. Such an approach maps the time series retrieval into a sequence retrieval problem. The fact that it can discern two data sets with similar auto-correlation function establishes its sensitivity to higher order statistics. A more detailed study is necessary for practical implementation of the proposed approach.

## 6. Acknowledgements

This research was supported by Symbram LLC, CA.